\documentstyle[preprint,aps]{revtex}
\begin{document}
\draft
\title{Mean Field Theory for Lossy Nonlinear Composites}
\author{K. W. Yu}
\address{Department of Physics, The Chinese University of Hong Kong, \\
         Shatin, New Territories, Hong Kong}
\maketitle

\begin{abstract}
The mean-field theory for lossy nonlinear composites, described by
complex and field-dependent dielectric functions, is presented.
By using the spectral representation of linear composites with identical
microstructure, we develop self-consistent equations for the effective
response. We examine two types of microstructure, namely, the
Maxwell-Garnett approximation and the effective medium approximation
to illustrate the theory.
\end{abstract}
\vskip 5mm

\section{Introduction}

Recently, we proposed a mean-field theory (MFT) for nonlinear composites
\cite{MFT}. The theory was applied to nonlinear transport properties of
conducting composites at zero frequency, described by real conductivities
and provided good agreement with numerical simulation \cite{MFT}.
However, for dielectric materials near relaxation, the loss may be
important.
It is tempting to extend the MFT to complex and nonlinear permitivities.
The local constitutive relation of the composite system is given by
\begin{eqnarray}
{\bf D}=\chi ({\bf E}^* \cdot {\bf E}) {\bf E},
\label{DE1}
\end{eqnarray}
where $\chi=\chi({\bf x})$ is the (position dependent) complex
third-order nonlinear coefficient. In what follows, we denote
$|{\bf E}|^2={\bf E}^* \cdot {\bf E}$ and
${\bf E}^2={\bf E} \cdot {\bf E}$ for convenience.
The effective response is defined as the volume average of {\bf D}
\begin{eqnarray}
\chi_e {\bf E}_0^2 {\bf E}_0={1\over V}\int_V\ dV\ {\bf D}({\bf x}),
\end{eqnarray}
where $V$ is the volume of composites and ${\bf E}_0$ is the applied
(average) field and is taken to be real:
\begin{eqnarray}
{\bf E}_0={1\over V}\int_V\ dV\ {\bf E}({\bf x}).
\end{eqnarray}
The corresponding boundary-value problem of nonlinear composite media
consists of the field equations:
\begin{eqnarray}
\nabla \cdot {\bf D} = 0, \\
\nabla \times {\bf E} = 0.
\end{eqnarray}
There exists a potential $\varphi$ so that ${\bf E}=\nabla \varphi$.
The boundary condition for $\varphi$ is
\begin{eqnarray}
\varphi=\varphi_0={\bf E}_0 \cdot {\bf x}
\end{eqnarray}
on surface ($S$) of the composite and $\varphi_0$ is real.
Consider the integral $(1/V)\int_V\ dV\ {\bf D}\cdot {\bf E}$. Through the
field equations, we can convert it into an integral over the surface ($S$)
of the composite:
$(1/V)\int_S\ d{\bf S} \cdot ({\bf D}\varphi)$, which is equal to
$(1/V)\int_S\ d{\bf S} \cdot ({\bf D}\varphi_0)$ because
$\varphi=\varphi_0$ on $S$. Converting the surface integral back to
the volume integral, we obtain:
\begin{eqnarray}
{1\over V}\int_V\ dV\ {\bf D}\cdot {\bf E}
&=& {1\over V}\int_V\ dV\ {\bf D}\cdot {\bf E}_0.
\label{W1}
\end{eqnarray}
Hence we find an equivalent expression for the effective response:
\begin{eqnarray}
\chi_e {\bf E}_0^4={1\over V}\int_V\ dV\ {\bf D} \cdot {\bf E}.
\label{8}
\end{eqnarray}
Let us consider a lossy nonlinear composite in which inclusions of complex
nonlinear coefficient $\chi_1$, present at volume fraction $p_1$, are
randomly embedded in a host medium of $\chi_2$, present at volume fraction
$p_2$, with the application of an external uniform field ${\bf E}_0$.
Note that $p_1+p_2=1$. From Eqs.(\ref{DE1}) and (\ref{8}), we have
therefore an expression for the effective response:
\begin{eqnarray}
\chi_e {\bf E}_0^4=p_1\chi_1 \langle |{\bf E}_1|^2 {\bf E}_1^2 \rangle +
                   p_2\chi_2 \langle |{\bf E}_2|^2 {\bf E}_2^2 \rangle,
\label{xe1}
\end{eqnarray}
where
\begin{eqnarray}
\langle |{\bf E}_i|^2 {\bf E}_i^2 \rangle = {1\over V_i}\int_{V_i}\ dV\
                   |{\bf E}|^2 {\bf E}^2, \hspace{1cm} i=1,2.
\label{AE2E2}
\end{eqnarray}
denotes the local field average within the $i$th component.
Similar to the derivation of Eq.(\ref{W1}), as $\varphi^*=\varphi_0$ on
$S$, we arrive at
\begin{eqnarray}
{1\over V}\int_V\ dV\ {\bf D}\cdot {\bf E}^* =
{1\over V}\int_V\ dV\ {\bf D}\cdot {\bf E}_0,
\label{W2}
\end{eqnarray}
which implies an alternative expression for the effective response:
\begin{eqnarray}
\chi_e {\bf E}_0^4=p_1\chi_1 \langle (|{\bf E}_1|^2)^2\rangle +
                   p_2\chi_2 \langle (|{\bf E}_2|^2)^2\rangle,
\label{xe2}
\end{eqnarray}
where
\begin{eqnarray}
\langle (|{\bf E}_i|^2)^2 \rangle = {1\over V_i}\int_{V_i}\ dV\
                   (|{\bf E}|^2)^2, \hspace{1cm} i=1,2.
\label{AE22}
\end{eqnarray}
We should remark that the different types of local field averages defined
in Eqs.(\ref{AE2E2}) and (\ref{AE22}) are generally not equal to each other.
It is interesting to note that they both give the same effective response
[Eqs.(\ref{xe1}) and (\ref{xe2})].
However, the nonlinear partial differential equations pertaining to the
boundary-value problem cannot be solved analytically. We shall invoke the
mean-field theory \cite{MFT} to obtain an approximate expression for the
effective response. According to previous work \cite{MFT}, we consider
the linear constitutive relation:
\begin{eqnarray}
{\bf D}=\epsilon {\bf E},
\end{eqnarray}
where $\epsilon=\epsilon({\bf x})$ is the (position dependent) complex
dielectric constant. By definition, we find the effective linear response:
\begin{eqnarray}
\epsilon_e {\bf E}_0={1\over V}\int_V\ dV\ \epsilon {\bf E}.
\end{eqnarray}
From Eq.(\ref{W1}), we have therefore
\begin{eqnarray}
\epsilon_e {\bf E}_0^2=p_1\epsilon_1 \langle {\bf E}_1^2 \rangle +
                       p_2\epsilon_2 \langle {\bf E}_2^2 \rangle.
\label{e0}
\end{eqnarray}
From Eq.(\ref{W2}), we have alternatively
\begin{eqnarray}
\epsilon_e {\bf E}_0^2=p_1\epsilon_1 \langle |{\bf E}_1|^2 \rangle +
                       p_2\epsilon_2 \langle |{\bf E}_2|^2 \rangle.
\label{e1}
\end{eqnarray}
This expression will be useful for calculating
$\langle |{\bf E}_i|^2 \rangle$.
In order to estimate $\chi_e$, we invoke the decoupling approximation
\cite{dca}
\begin{eqnarray}
\langle |{\bf E}_i|^2 {\bf E}_i^2 \rangle =
\langle |{\bf E}_i|^2 \rangle \langle {\bf E}_i^2 \rangle,
\hspace{1cm} i=1,2.
\end{eqnarray}
within the $i$th component. This assumption is good in microstructure
for which the local electric field is nearly uniform within the $i$th
component, but less accurate when these fluctuations are large, as in a
random composite near the percolation threshold \cite{dca}.
With this assumption,
\begin{eqnarray}
\chi_e {\bf E}_0^4=p_1\chi_1 \langle |{\bf E}_1|^2 \rangle
                             \langle {\bf E}_1^2 \rangle +
                   p_2\chi_2 \langle |{\bf E}_2|^2 \rangle
                             \langle {\bf E}_2^2 \rangle.
\label{e2}
\end{eqnarray}
Comparing Eqs.(\ref{e0}) and (\ref{e2}), it is tempting to write
$\epsilon_1=\chi_1 \langle |{\bf E}_1|^2 \rangle$,
$\epsilon_2=\chi_2 \langle |{\bf E}_2|^2 \rangle$ and
$\epsilon_e=\chi_e {\bf E}_0^2$.
Hence, within the decoupling approximation, we may interpret the nonlinear
component as a linear dielectric material with a field-dependent dielectric
constant. Suppose $\epsilon_e$ is known as a function of its constituent
dielectric constants,
\begin{eqnarray}
\epsilon_e=F(\epsilon_1, \epsilon_2, p_1),
\end{eqnarray}
then we find the local field averages \cite{MFT}
\begin{eqnarray}
p_1\langle {\bf E}_1^2 \rangle={\partial \epsilon_e \over
                                \partial \epsilon_1}{\bf E}_0^2,
\label{p1E1} \\
p_2\langle {\bf E}_2^2 \rangle={\partial \epsilon_e \over
                                \partial \epsilon_2}{\bf E}_0^2.
\label{p2E2}
\end{eqnarray}
However, we have to determine $\langle |{\bf E}_i|^2 \rangle$ in the
$i$th component, which is generally not equal to
$\langle {\bf E}_i^2 \rangle$.
Thus a straightforward application of the previous formalism \cite{MFT}
is impossible. We resort to an alternative approach based on the spectral
representation \cite{spectral}.

In what follows, we shall consider two important types of microstructures:
(1) Dispersion microstructures as in the Maxwell-Garnett approximation
\cite{MGA} and (2) symmetric microstructures as in the Bruggeman
effective medium approximation \cite{EMA}.

\section{Maxwell-Garnett Approximation}

The Maxwell-Garnett approximation (MGA) is good for dispersion
microstructures and the theory is inherently non-symmetrical \cite{MGA}.
For convenience, we consider the case in which component 1 is embedded
in component 2.
For a linear composite of $\epsilon_1$ and $\epsilon_2$ at volume
fractions $p_1$ and $p_2$ respectively, the MGA reads \cite{MGA}:
\begin{eqnarray}
{\epsilon_e-\epsilon_2 \over \epsilon_e+(d-1)\epsilon_2} =
p_1{\epsilon_1-\epsilon_2 \over \epsilon_1+(d-1)\epsilon_2},
\label{MGA}
\end{eqnarray}
where $d$ is the dimensionality of composites.
Solving Eq.(\ref{MGA}), we obtain
\begin{eqnarray}
\epsilon_e={\epsilon_2 [\epsilon_1(dp_1+p_2)+\epsilon_2(d-1)p_2] \over
            p_2\epsilon_1+(d-p_2)\epsilon_2}.
\label{MGAe}
\end{eqnarray}
From Eqs.(\ref{p1E1}) and (\ref{p2E2}), we can calculate the local field
averages:
\begin{eqnarray}
p_1 \langle {\bf E}_1^2 \rangle = {p_1 d^2 \epsilon_2^2 \over
            [p_2\epsilon_1+(d-p_2)\epsilon_2]^2} {\bf E}_0^2.
\label{local}
\end{eqnarray}
\begin{eqnarray}
p_2 \langle {\bf E}_2^2 \rangle =\left(1 -
           {p_1 d [d\epsilon_2^2 - p_2(\epsilon_1-\epsilon_2)^2] \over
            [p_2\epsilon_1+(d-p_2)\epsilon_2]^2} \right) {\bf E}_0^2.
\end{eqnarray}
In MGA, the local field in the inclusion is uniform, we can determine
${\bf E}_1$ and hence ${\bf E}_1^*$ explicitly, we find
\begin{eqnarray}
p_1 \langle |{\bf E}_1|^2 \rangle = {p_1 d^2 |\epsilon_2|^2 \over
            |p_2\epsilon_1+(d-p_2)\epsilon_2|^2} {\bf E}_0^2.
\label{MGA1}
\end{eqnarray}
From Eq.(\ref{e1}) and using Eqs.(\ref{MGAe}) and (\ref{MGA1}), we find
\begin{eqnarray}
p_2 \langle |{\bf E}_2|^2 \rangle =\left(1 -
           {p_1 d [d|\epsilon_2|^2 - p_2|\epsilon_1-\epsilon_2|^2] \over
            |p_2\epsilon_1+(d-p_2)\epsilon_2|^2} \right) {\bf E}_0^2.
\label{MGA2}
\end{eqnarray}
We should remark that although
$|\langle {\bf E}_1^2 \rangle|=\langle |{\bf E}_1|^2 \rangle$ in MGA,
$|\langle {\bf E}_2^2 \rangle|$ is generally not equal to
$\langle |{\bf E}_2|^2 \rangle$. In fact,
$|\langle {\bf E}_2^2 \rangle| \le \langle |{\bf E}_2|^2 \rangle$.
A common error in the literature is to treat these averages equal even
when $\chi$ is complex.
If we write $\epsilon_1=\chi_1 \langle |{\bf E}_1|^2 \rangle$
and $\epsilon_2=\chi_2 \langle |{\bf E}_2|^2 \rangle$, then 
Eqs.(\ref{MGA1})--(\ref{MGA2})
can be solved self-consistently for $\langle |{\bf E}_1|^2 \rangle$
and $\langle |{\bf E}_2|^2 \rangle$ and hence the effective nonlinear
response can be calculated.

In Fig.\ref{fig1}, we present the MGA results in three dimensions (3D).
We let $\chi_1=1+3i$ and $\chi_2=3+i$ and compute $\chi_e$ as a function
of $p_1$.
As $p_1$ increases, Re($\chi_e$) decreases from Re($\chi_2)=3$ towards
Re($\chi_1)=1$ while Im($\chi_e$) increases from Im($\chi_2)=1$ towards
Im($\chi_1)=3$. Since the contrast between the two components is relatively
small, the local field averages $\langle |{\bf E}_1|^2 \rangle$ and
$\langle |{\bf E}_2|^2 \rangle$ remain close to unity. However, unlike
$|\langle {\bf E}_1^2 \rangle|$, $|\langle {\bf E}_2^2 \rangle|$ shows
a significant deviation from $\langle |{\bf E}_2|^2 \rangle$, indicating
the nonuniformity of local field in the host.

For two-component composites, it has proved convenient to adopt the
spectral representation of the effective linear response \cite{spectral}:
Let $v=1-\epsilon_1/\epsilon_2$, $w=1-\epsilon_e/\epsilon_2$, and $s=1/v$,
we find
\begin{eqnarray}
w(s)=\int_0^1 {m(s'){\rm d}s' \over s - s'},
\end{eqnarray}
where $m(s)$ is the spectral density which is obtained through a limiting
process:
\begin{eqnarray}
m(s) = \lim_{\eta \to 0^+} -{1\over \pi}{\rm Im}\ w(s+i\eta).
\label{ms}
\end{eqnarray}
Furthermore, $m(s)$ obeys the sum rule
\begin{eqnarray}
\int_0^1 m(s'){\rm d}s' = p_1.
\end{eqnarray}
Eqs.(\ref{MGAe})--(\ref{MGA2}) can readily be converted into the spectral
representation. We find
\begin{eqnarray}
w={p_1\over s-s_1},
\end{eqnarray}
where $s_1=p_2/d$. From Eq.(\ref{ms}), we obtain the spectral density in
MGA:
\begin{eqnarray}
m(s)=p_1 \delta(s-s_1).
\label{ms1}
\end{eqnarray}
Similarly, the local field averages [Eqs.(\ref{local})--(\ref{MGA2})]
are given by
\begin{eqnarray}
\langle {\bf E}_1^2 \rangle = { {\bf E}_0^2 \over (1-vs_1)^2},
\label{MGAE1} \\
\langle |{\bf E}_1|^2 \rangle = { {\bf E}_0^2 \over |1-vs_1|^2},
\label{MGAAE1}
\end{eqnarray}
\begin{eqnarray}
p_2\langle {\bf E}_2^2 \rangle
=\left( 1- {p_1 (1-v^2s_1) \over (1-vs_1)^2}\right) {\bf E}_0^2,
\label{MGAE2} \\
p_2\langle |{\bf E}_2|^2 \rangle
=\left( 1- {p_1 (1-|v|^2s_1) \over |1-vs_1|^2}\right) {\bf E}_0^2.
\label{MGAAE2}
\end{eqnarray}
If we let $v=1-\chi_1 \langle |{\bf E}_1|^2 \rangle/\chi_2
                      \langle |{\bf E}_2|^2 \rangle$
and solve Eqs.(\ref{MGAAE1}) and (\ref{MGAAE2}) self-consistently for
$\langle |{\bf E}_1|^2 \rangle$ and $\langle |{\bf E}_2|^2 \rangle$,
the effective nonlinear response can be determined. Although the
solution appears somewhat simpler, the same results are obtained.

\section{Effective Medium Approximation}

The Bruggeman effective-medium approximation (EMA) is known to be
symmetrical with respect to interchanging components 1 and 2 \cite{EMA}.
Note that there is a percolation threshold in the theory. For a random
linear composite of $\epsilon_1$ and $\epsilon_2$ at volume fractions
$p_1$ and $p_2$ respectively, the self-consistency equations reads
\cite{EMA}:
\begin{eqnarray}
p_1{\epsilon_1-\epsilon_e \over \epsilon_1+(d-1)\epsilon_e} +
p_2{\epsilon_2-\epsilon_e \over \epsilon_2+(d-1)\epsilon_e} = 0,
\label{EMA}
\end{eqnarray}
where $d$ is the dimensionality of composites.
Solving Eq.(\ref{EMA}), we obtain $\epsilon_e$ and hence $w$ and the
spectral density
\begin{eqnarray}
m(s)={dp_1-1\over d-1}\delta(s)\theta(dp_1-1)+m_1(s),
\end{eqnarray}
where $\delta(s)$ and $\theta(dp_1-1)$ denote the Dirac delta function
and the Heavyside step function respectively, and
\begin{eqnarray}
m_1(s) &=& {d\over 2\pi (d-1)s}\sqrt{(s-s_1)(s_2-s)}, \hspace{1cm}
           s_1 < s < s_2, \\ \nonumber
       &=& 0, \hspace{1cm} {\rm otherwise},
\end{eqnarray}
where $s_1$ and $s_2$ are given by
\begin{eqnarray}
s_1={1\over d}\left( 1+(d-2)p_1-2\sqrt{(d-1)p_1(1-p_1)} \right),
     \nonumber \\
s_2={1\over d}\left( 1+(d-2)p_1+2\sqrt{(d-1)p_1(1-p_1)} \right).
\end{eqnarray}
When $p_1>1/d$, which is the percolation threshold in EMA, component 1
percolates the composite and there is a $\delta$ function contribution
to $m(s)$ at $s=0$. Eqs.(\ref{MGAE1})--(\ref{MGAAE2}) can readily be
generalized to \cite{Sheng}:
\begin{eqnarray}
p_1\langle {\bf E}_1^2 \rangle=
   \int_0^1  {m(s'){\rm d}s' \over (1-vs')^2}{\bf E}_0^2, \\
p_1\langle |{\bf E}_1|^2 \rangle=
   \int_0^1  {m(s'){\rm d}s' \over |1-vs'|^2}{\bf E}_0^2,
\label{EMAAE1}
\end{eqnarray}
\begin{eqnarray}
p_2\langle {\bf E}_2^2 \rangle=
\left( 1- \int_0^1 {(1-v^2s')m(s'){\rm d}s' \over (1-vs')^2}\right)
   {\bf E}_0^2, \\
p_2\langle |{\bf E}_2|^2 \rangle=
\left( 1- \int_0^1 {(1-|v|^2s')m(s'){\rm d}s' \over |1-vs'|^2}\right)
   {\bf E}_0^2.
\label{EMAAE2}
\end{eqnarray}
One can check that when the MGA spectral density [Eq.(\ref{ms1})] is used,
Eqs.(\ref{MGAE1})--(\ref{MGAAE2}) are recovered.
Again we let $v=1-\chi_1 \langle |{\bf E}_1|^2 \rangle/\chi_2
\langle |{\bf E}_2|^2 \rangle$. Eqs.(\ref{EMAAE1}) and (\ref{EMAAE2})
are coupled integral equations. They can be solved numerically for
$\langle |{\bf E}_1|^2 \rangle$ and $\langle |{\bf E}_2|^2 \rangle$
and hence the effective nonlinear response can be determined.

In Fig.\ref{fig2}, we present the EMA results in 3D. We let $\chi_1=1+3i$
and $\chi_2=3+i$. Fig.\ref{fig2} exhibits similar behavior as Fig.1,
however, with
some differences. Here Re($\chi_e$) and Im($\chi_e$) cross at $p_1=0.5$
because EMA describes the symmetric microstructure and we have the
symmetric choice of $\chi_1$ and $\chi_2$. Other quantities, like the
local field averages
$\langle |{\bf E}_1|^2 \rangle$, $\langle |{\bf E}_2|^2 \rangle$, and
$|\langle {\bf E}_1^2 \rangle|$, $|\langle {\bf E}_2^2 \rangle|$
also cross at $p_1=0.5$. In EMA, both the local fields in the inclusion
and host regions are not uniform.

\section*{Discussion and Conclusion}
The model may be applicable to lossy dielectric composite materials under
an intense electric, available from laser.
The present formalism can naturally be extended to optical properties of
nonlinear composites \cite{optical}. When the size of the inclusion is
smaller than the wavelength of light, we can use the quasi-static limit.
The mean-field theory can be used to calculate the optical nonlinearity.

In conclusion, we extend the mean-field theory for lossy nonlinear
composites, described by complex and field-dependent dielectric functions.
We developed self-consistent equations with the aid of the spectral
representation of the linear composites.
We examine two important types of microstructure, namely,
the Maxwell-Garnett approximation and the effective medium approximation
to illustrate the present theory.

\section*{Acknowledgments}
This work was supported by the Research Grants Council under project
number CUHK 461/95P. I acknowledge useful conversation with
Professor Ping Sheng.

\begin{figure}
\caption{Real and imaginary parts of the effective nonlinear response
plotted against volume fraction of the inclusion in the MGA with
$\chi_1=1+3i$ and $\chi_2=3+i$.
\label{fig1}}

\vskip 5mm
\caption{Real and imaginary parts of the effective nonlinear response
plotted against volume fraction of the inclusion in the EMA with
$\chi_1=1+3i$ and $\chi_2=3+i$.
\label{fig2}}
\end{figure}

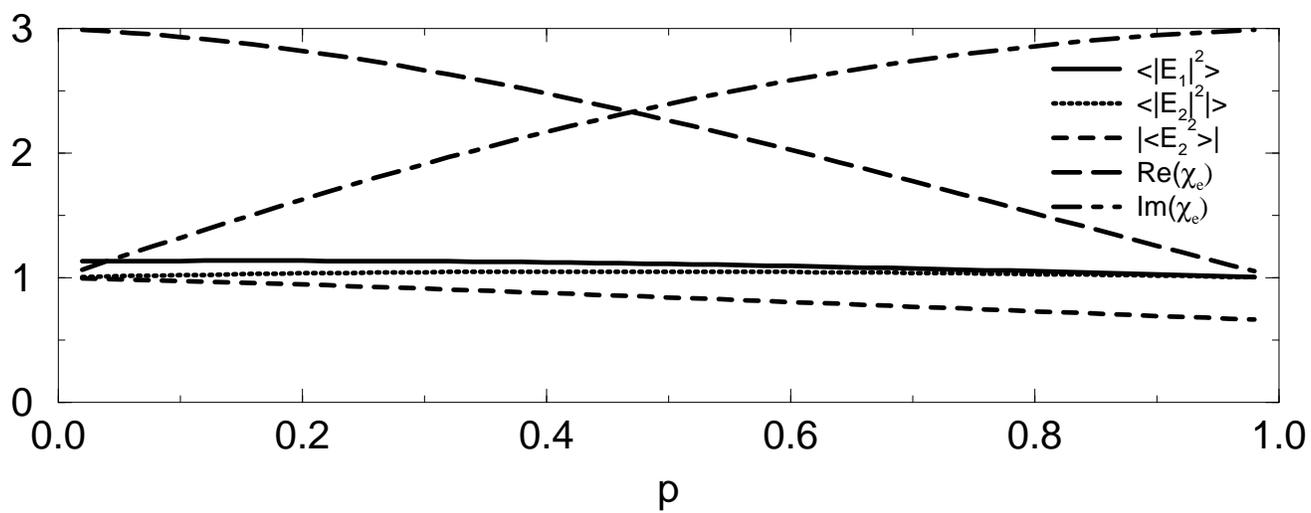
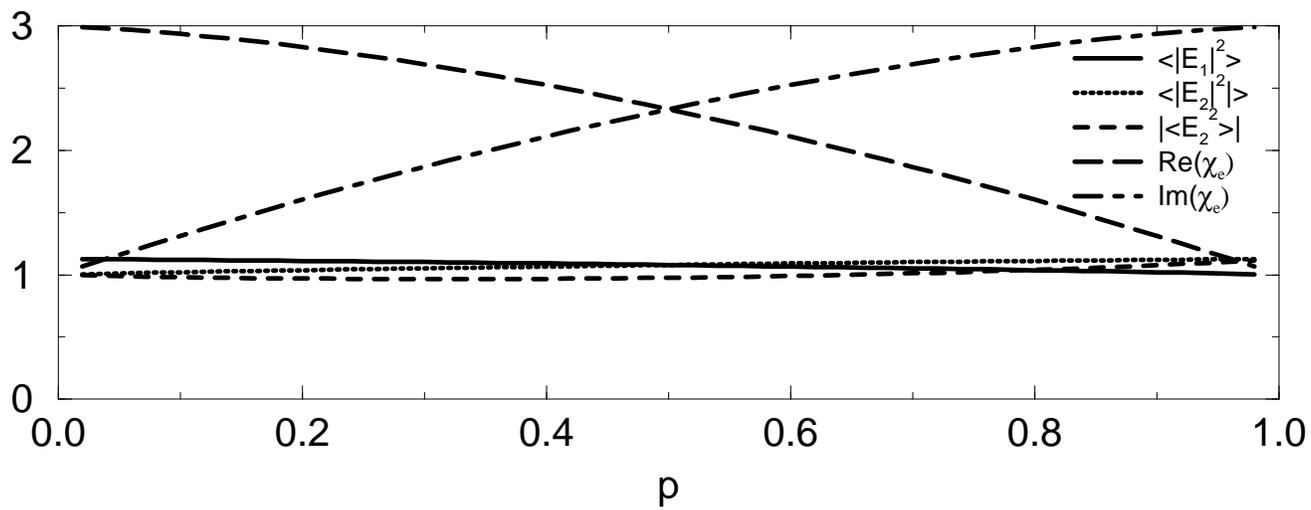
\end{document}